\documentclass[journal=jacsat,manuscript=article]{achemso}

\usepackage{chemformula} 
\usepackage[T1]{fontenc} 
\usepackage{graphicx}
\usepackage{epstopdf}
\usepackage{xcolor}

\author{Giovanna Palermo}
\affiliation{Department of Physics, University of Calabria, Via P. Bucci, 87036 Rende (CS), Italy}
\alsoaffiliation{CNR NANOTEC - Istituto di Nanotecnologia, UOS Cosenza, 87036, Rende (CS), Italy}
\email{giovanna.palermo@unical.it}
\author{Giuseppe E. Lio}
\affiliation{Department of Physics, University of Calabria, Via P. Bucci, 87036 Rende (CS), Italy}
\alsoaffiliation{CNR NANOTEC - Istituto di Nanotecnologia, UOS Cosenza, 87036, Rende (CS), Italy}
\author{Marco Esposito}
\affiliation{CNR NANOTEC- Istituto di Nanotecnologia, Polo di Nanotecnologia, Campus Ecotekne, Lecce, Italy}
\author{Loredana Ricciardi}
\affiliation{CNR NANOTEC - Istituto di Nanotecnologia, UOS Cosenza, 87036, Rende (CS), Italy}
\author{Mariachiara Manoccio}
\affiliation{CNR NANOTEC- Istituto di Nanotecnologia, Polo di Nanotecnologia, Campus Ecotekne, Lecce, Italy}
\author{Vittorianna Tasco}
\affiliation{CNR NANOTEC- Istituto di Nanotecnologia, Polo di Nanotecnologia, Campus Ecotekne, Lecce, Italy}
\author{Adriana Passaseo}
\affiliation{CNR NANOTEC- Istituto di Nanotecnologia, Polo di Nanotecnologia, Campus Ecotekne, Lecce, Italy}
\author{Antonio De Luca}
\affiliation{Department of Physics, University of Calabria, Via P. Bucci, 87036 Rende (CS), Italy}
\alsoaffiliation{CNR NANOTEC - Istituto di Nanotecnologia, UOS Cosenza, 87036, Rende (CS), Italy}
\email{antonio.deluca@unical.it}
\author{Giuseppe Strangi}
\affiliation{Department of Physics, University of Calabria, Via P. Bucci, 87036 Rende (CS), Italy}
\alsoaffiliation{CNR NANOTEC - Istituto di Nanotecnologia, UOS Cosenza, 87036, Rende (CS), Italy}
\alsoaffiliation[Second University]
{Department of Physics, Case Western Reserve University, 10600 Euclid Avenue, Cleveland, Ohio 44106, USA}
\email{gxs284@case.edu}

\title[An \textsf{achemso} demo]
  {Biomolecular sensing at the interface between chiral metasurfaces and hyperbolic metamaterials}

\begin{document}

\begin{abstract}

In recent years significant efforts have been made to design and fabricate functional nanomaterials for biomedical applications  based on the control of light matter interaction at the nanometer scale.
Among many other artificial materials, hyperbolic dispersion metamaterials allow to access unprecedented physical effects and mechanisms  due to the extreme anisotropy of their optical constants.
The unbound isofrequency surface of hyperbolic metamaterials (HMMs) enable the possibility to support a virtually infinite density of states and ultra-high confinement of electromagnetic fields, allowing perfect absorption of light and extreme sensing properties. 
Optical sensor technology based on plasmonic metamaterials offers significant opportunities in the field of clinical diagnostics, particularly for the detection of low-molecular-weight biomolecules in highly diluted solutions. 
In this context, we present a  computational effort to engineer a biosensing platform based on hyperbolic metamaterials, supporting highly confined bulk plasmon modes integrated with out-of-plane chiral metasurfaces. The role of the helicoidal chiral metasurface is manifold:
i) as a diffractive element to increase the momentum of the incoming light to excite the plasmon sensing modes with linearly and circularly polarized light; ii) as out-of-plane extended sensing surface to capture target analytes away from the substrate thereby the diffusion limit; iii) as a plamonic chiral nanostructure with enhanced sensing performance over circularly polarized reflectance light. 
\end{abstract}


\section{}

In the fight against cancer it is essential to diagnose the disease at an early stage. To date, one of the most widely used methods for the diagnosis and staging of the disease is represented by the tissue biopsy.  However, this clinical exam is invasive and it is performed when the site of the tumor has already been identified. 
In contrast, liquid biopsy has proved to be a valid alternative to detect and monitor cancer biomarkers. \cite{joosse2013biologic, palmirotta2018liquid, cohen2018detection, gorgannezhad2018circulating, mattox2019applications, lim2019liquid}
For this reason, researchers are focused to develop  extremely sensitive platforms, able to determine specific nucleic acids and proteins in circulating tumor cells (CTCs), in body fluids as  blood, urine, sweat and tears.

To the present, several investigation techniques have been developed to identify for example the presence of proteins in the blood by means of an electrochemical analysis,\cite{leca2005biosensors} or in saliva by means of electrochemical immunoassay biosensors. \cite{corrie2015blood}

Optical sensor technology based on plasmonic metamaterials offers significant opportunities in the field of clinical diagnosis, particularly for the detection of low-molecular weight biomolecules in highly diluted solutions.\cite{sreekanth2016extreme,baqir2018nanoscale,sreekanth2016biosensing}

Hyperbolic metamaterials (HMMs) represent a novel class of promising artificial plasmonic materials used in the field of bio-sensing, supporting highly confined bulk plasmon polaritons (BPPs) in addition to surface plasmon polaritons (SPPs).\cite{poddubny2013hyperbolic,ferrari2015hyperbolic,shekhar2014strong, shekhar2014hyperbolic}

It has been experimentally demonstrated that  it is possible to excite both surface and bulk plasmon polaritons in HMM by means of a hypergrating, consisting of 1D and 2D metallic diffraction gratings, such as stripes, disk or holes gratings.\cite{sreekanth2013experimental} However, further sensing properties could be reached by engineering the hypergrating features towards the realization of a high sensitivity and high accuracy biosensing platform.

Here, we report, for the first time, a numerical simulation of a new biosensing platform based on HMMs integrated with an out-of-plane 3D chiral metasurface hypergrating (CMH), where the fundamental building block is represented by a helical shaped nanostructure. 
In particular, the implementation of 3D metallic grating could enable the excitation of the high-k modes of the underlying HMM and reduce the diffusion limit by the detection of molecules that do not necessarily flows close to the surface. Moreover, the strong intrinsic chirality belonging to plasmonic nanohelices induces circular polarization dependent BPP mode excitation, resulting in a giant circular dichroism value (CD), boosting the sensing properties of the system. 

We demonstrate that our advanced biosensing platform exhibits extremely high sensing properties since it takes advantages of the elevated sensitivity of the HMM bulk modes (high Q factor) and the significant accuracy of the circular dichroism (CD) signal of the CMH that exhibits more spectral features to track the environment changes. 

Moreover, considering the increased out-of-plane surface of the 3D nanohelices, able to that interact with analyte targets, we have introduced the mass sensitivity as analytical figure of merit, that describes the surface coverage-refractive index variation relationship. We have calculated that the minimum quantity of binding analyte that can leads to an appreciable variation in terms of refractive index unit (RIU) results equal to 0.04 pg/mm$^2$. 

Similarly, we carried out a study of the analyte diffusion calculating the sensitivity changes as a function of the relative surface coverage, at different distance away from the HMM surface, noting that our system allows to detect analytes over the diffusion limit of planar hypergrating.

Finally, we quantified the sensitivity and Figure of merit (FOM) of the sensor tracking the spectral shift of the CD features as a function of the glycerol/water concentration ratio leading to a remarkable FOM value of about 20.0 RIU$^{-1}$.
 
These results indicate that our innovative sensing platform offers unique advantages for high sensitivity detection of analytes in many biological applications.

 
\section{Results and discussion}

\begin{figure}[!h]
  \includegraphics[width=0.7\columnwidth]{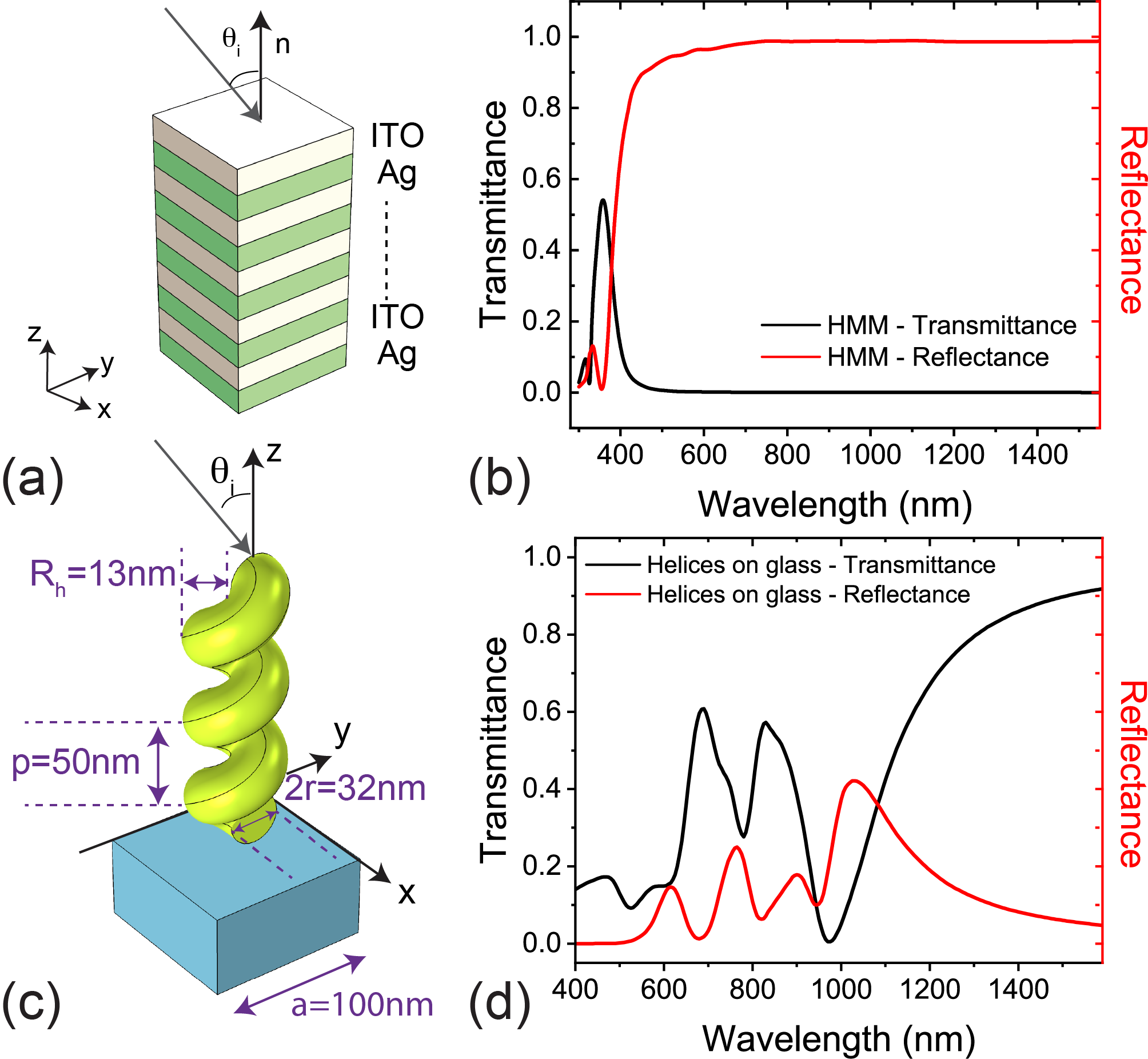}
  \caption{(a) Schematic diagram of the ITO/Ag HMM and (b) corresponding calculated Reflectance and Transmittance spectra. (c) Sketch of one lateral unit cell of the out-of-plane chiral structure composed of a right-handend Au helix on a glass substrate, the relevant structure parameters are illustrated. (d) Calculated Reflectance and Transmittance spectra for the chiral metasurface.  The reflectance and transmittance spectra are calculated for TM wave, angle of incidence $\theta_i =50^\circ$.}
  \label{fig:1}
\end{figure}

The sketch reported in Figure \ref{fig:1}a shows the schematic diagram of a type II HMM configuration consisting of alternating dielectric and metallic thin films. In particular, to obtain a hyperbolic dispersion at optical frequencies (with dielectric permittivity tensor component $\varepsilon_{//} < 0$ and $\varepsilon_{\perp} > 0$, above 418 nm), we designed a HMM composed by indium tin oxide (ITO - 20 nm) and silver (Ag - 20 nm) layers using Effective Medium Theory (EMT)\cite{choy2015effective} (see Supporting Information).

Figure \ref{fig:1}b shows the numerical reflectance and transmittance curves of the HMM, provided by COMSOL Multiphysics simulation.\cite{multiphysics1998introduction}
The reflectance and transmittance are calculated by considering a TM-wave (p-polarization) and an incident angle $\theta_i =50^\circ$. 
The minimum value at about 358 nm in the reflectance curve (and the related maximum in the transmittance one) is referred to as the Ferrel-Berreman mode for silver nanometric layers.\cite{caligiuri2016dielectric}

In order to excite both surface and bulk plasmon modes inside the HMM, we designed an out-of-plane CMH as shown in Figure \ref{fig:1}c.
The geometrical parameters of the helix array are wire radius ($r$), helix radius ($R_h$), pitch height or axial pitch ($p$), the number of pitches ($N_p$) and the lattice constant ($a$).\cite{gansel2010gold}
Currently CMH with these geometrical features could be fabricated by focused ion beam induced deposition (FIBID),\cite{esposito2014three} or by the shadow-growth or glancing-angle deposition techniques. \cite{mark2013hybrid, hawkeye2007glancing, robbie1996chiral, zhao2003designing}

The simulations were conducted considering an infinite array of helices both on a glass substrate and onto the HMM structures. This has been possible by simulating a single unit cell of the chiral metasurface composed of a right-handed Au helix arranged in a square lattice. 
In the first analysis distilled water (n = 1.3330) was considered as the surrounding medium for the chiral metasurface.
The reflectance and transmittance curves are calculated for a TM wave and same angle of incidence ($\theta_i =50^\circ$) with respect to the helix axis.
The resulting spectra, shown in Figure \ref{fig:1}d,  are determined by the interplay of internal resonances and their mutual coupling that strongly depends on the structure parameters of the CMH. 

\begin{figure}[!h]
  \includegraphics[width=0.8\columnwidth]{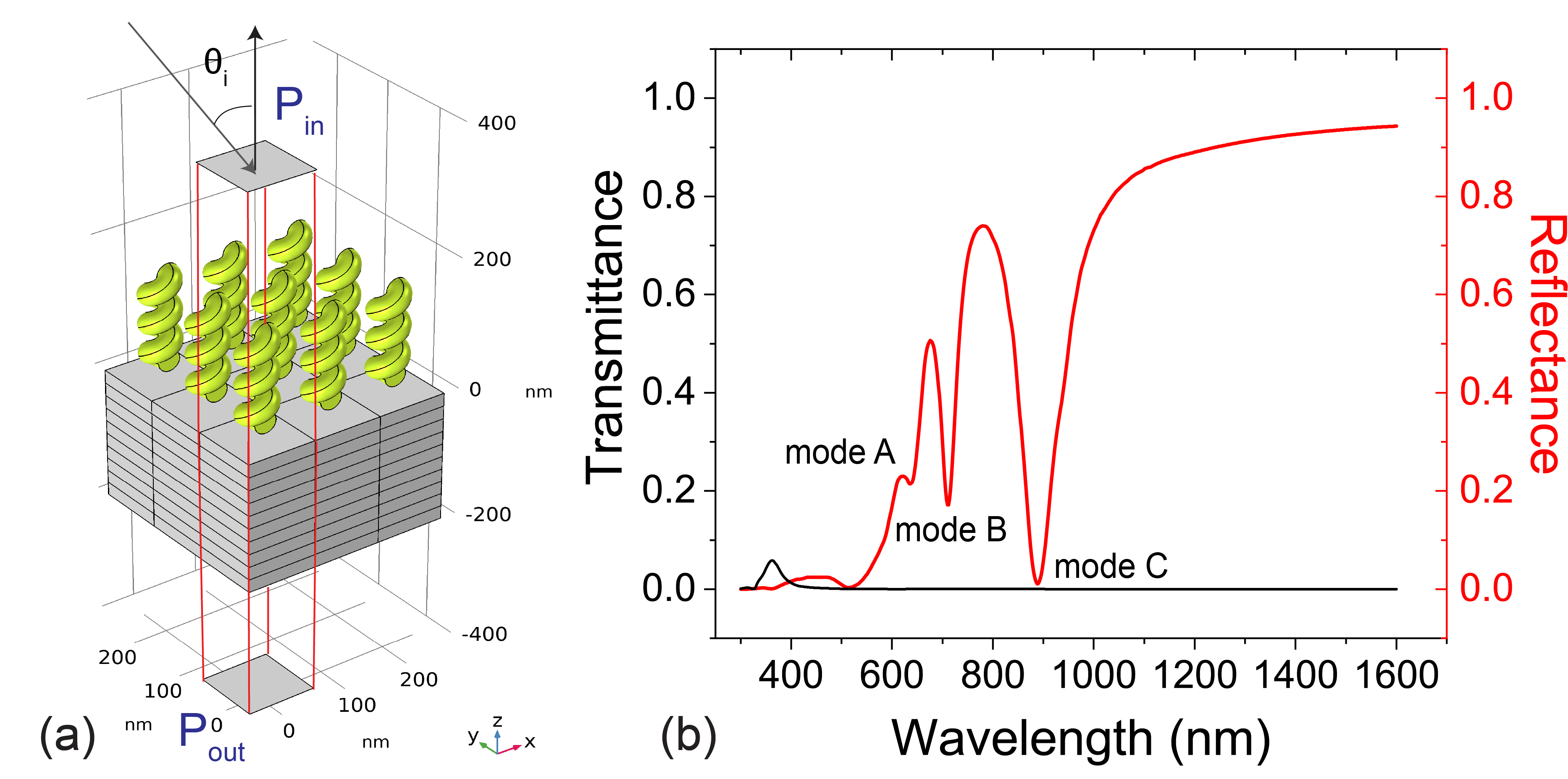}
  \caption{(a) Unit cell of the CMH-HMM simulated geometry. COMSOL permits to simulate sources or detectors of e.m. radiation by creating Ports: in our case, there is a port  on the top (P$_{in}$), from where the radiation propagates and a port on the bottom (P$_{out}$), that behaves as a detector. (b) Calculated Reflectance and Transmittance spectra of an Au helix array on HMM - water as surrounding medium - for TM, angle of incidence $\theta_i$=50$^\circ$.}
  \label{fig:2}
\end{figure}

It is important to note that the calculated reflectance for the chiral metasurface on glass substrate, does not exceed the 40\% in all the considered spectral range (400-1600 nm). 
In order to study the optical behavior in terms of reflectance and transmittance of the chiral metasurface on the HMM (CMH-HMM), we combined the two above mentioned structures into a single 3D geometry built in COMSOL.
The obtained 3D system  is composed by a parallelepiped 8 times higher than the lattice constant $a$ - Figure \ref{fig:2}a. The choice of the box height is fundamental for the optimal propagation of the lightwave inside the system preventing any diffraction or boundary problems. The numerical geometry of the model is sketched in Figure \ref{fig:2}a, with, starting from the top, the superstrate containing the Au helix (distilled water as surrounding medium), the multi-layer system (grey stack of ITO/Ag) and the substrate (glass).

Here, an electromagnetic plane wave impinges on the combined system from the superstrate, with a specific incident angle ($\theta_i$) and polarization (s-polarization (TE) or p-polarization (TM)). Light interacting with the structure is then collected through both its substrate (typically glass) and superstrate, allowing to compute the optical quantities of interest (transmittance (T) and reflectance (R)). 

This is done by solving the frequency-domain partial differential equation (PDE) that governs the $\mathbf{E}$ and $\mathbf{H}$ fields associated with the electromagnetic wave propagating through the structure:

\begin{equation}
\nabla \times\mu_r^{-1}(\nabla \times\mathbf{E})-\omega^2\varepsilon_0\mu_0(\varepsilon_r-i\sigma/\omega\varepsilon_0)\mathbf{E}=0
\label{eq1}
\end{equation}

In the condition of the electric conductivity $\sigma=0$ and non-magnetic materials ($\mu_r=1$), the previous equation reduces to:

\begin{equation}
\nabla \times(\nabla \times\mathbf{E})-k_0\textsuperscript{2}\varepsilon_r\mathbf{E}=0
\label{eq2}
\end{equation}

Here $k_0$ is the incident wavevector in vacuum or air ($k_0 = 2\pi/\lambda$), while $\varepsilon_r$ represents the material dielectric permittivity. By providing as an input the values of real and imaginary parts of the refractive index of any considered material, the software retrieves the corresponding dielectric permittivity ($\varepsilon_r$) and numerically solves Eq. \ref{eq2} to obtain the $\mathbf{E}$ field distribution.  Scattering parameters (S-parameter) for transmittance ($S_{21}$) and reflectance ($S_{11}$) are computed (See Supporting information).

In order to simulate the behavior of TM and TE  light in a 3D environment, it is necessary to properly write the components of the electromagnetic fields with respect to the incidence plane ($xz$) and set the polarizations accordingly: $\mathbf{H}= (0,1,0)$ for TM and  $\mathbf{E}=(0,1,0)$ for TE.\cite{lio2019comprehensive}

As we can see in Figure \ref{fig:2}b, the calculated transmittance for the CMH-HMM turns out to be zero in all the considered vis-NIR regions, while the reflectance spectrum, calculated for a TM and $\theta_i = 50^\circ$, is totally modified respect to the reflectance of the HMM and the Au helix array considered separately (see Figures 1b-1d). In particular, we can distinguish three reflectance minima at 635 nm (mode A), 710 nm (mode B) and 890 nm (mode C). 

It is important to note that these reflectance dips are closely related to the geometrical and material properties of the designed CMH. In fact, considering an out-of-plane 3D pillars hypergrating  characterized by the same sizes of the wire radius ($r$), or the helix radius ($R_h$) it does not leads to comparable signals usable for sensing (see Supplementary materials).

According to the grating coupling technique of surface plasmon excitation,\cite{sambles1991optical,sreekanth2013experimental} the surface plasmon modes can be excited when the wavevector of the grating diffraction orders are greater than that of the incident light. Under this condition, diffraction orders are no longer propagating waves, but evanescent field and the enhanced wavevector results to be responsible for the coupling of the incident light to the surface plasmon modes according to the coupling condition:\cite{barnes2004surface}

\begin{equation}
k^2_{SPP}=n^2_0 k^2_0 sin^2 \theta \pm 2 n_0 m k_g k_0 sin \theta cos \phi + (m k_g)^2
\label{eq3}
\end{equation}

where $n_0$ is the refractive index of the incident medium, $k_0$ is the vacuum wavevector, $m$ is the grating diffraction order and $\theta$ and $\phi$  are the incident grazing and azimuthal angle, respectively. $k_g = 2\pi/ \Lambda$ is the grating wavevector, with $\Lambda$ the grating period, in our case corresponding to the the lattice constant $a$. When $\phi =0$, Eq. \ref{eq3} results to be: $k_{SPP}=n_0 k_0 sin \theta \pm  m k_g$.  By using this equation we calculated the corresponding $k_x$ for our three modes: $k_{xA}$,$k_{xB}$ and $k_{xC}$.

\begin{figure}
  \includegraphics[width=0.7\columnwidth]{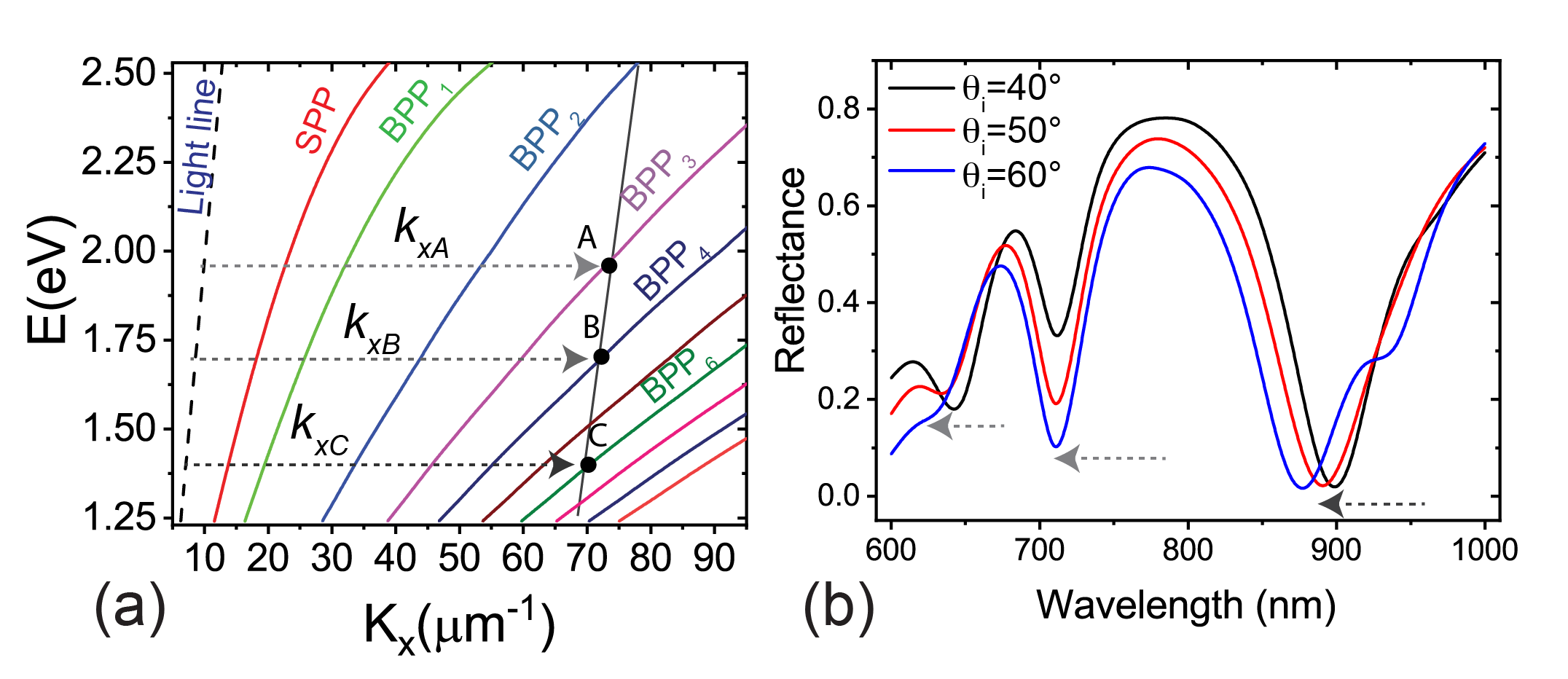}
  \caption{(a) Dispersion relations of SPP and BPP modes and (b) calculated reflection spectra for three different incident angles ($\theta_i$) for the CMH-HMM.}
  \label{fig:3}
\end{figure}

By considering the dispersion relations for the SPP and BPP modes given by:\cite{avrutsky2007highly}
\begin{equation}
k_{SPP}= k_0 \sqrt{\varepsilon_d  \varepsilon_m \over \varepsilon_d  + \varepsilon_m}
\label{eq4}
\end{equation}

\begin{equation}
k_{BPP_N}=k_0 \sqrt{\varepsilon_d  -  \frac{\lambda^2 N}{\pi^2 t_d t_m } \frac{\varepsilon_d }{\varepsilon_m}}
\label{eq5}
\end{equation}

we can  plot the SPP and BPP dispersion curves (Figure \ref{fig:3}a) supported by the structure. 
In the above equations, $t_d$, $\varepsilon_d$ and $t_m$, $\varepsilon_m$ are the thickness and dielectric permittivity of the dielectric and metal, in our case ITO and Ag, respectively, while $N$ represents the order mode ($N$=1,2,3...). 
In this plot it is possible to identify three points corresponding to the intersection between the $k_x$ component of modes A, B and C and the BPP modes of the structure, Figure \ref{fig:3}a.

Since both surface and bulk plasmon mode excitation depends on the incident angle, we calculated the reflectance of the CMH-HMM for different angle of incidence (Figure \ref{fig:3}b). We observed that there is a blue shift in the reflectance minima when the incident angle is increased. This is attributed to the variation in modal indices with incident angles.\cite{fan2006all}

The reported data clearly show that the CMH can excite the BPP modes of the underneath hyperbolic metamaterial. These modes are expected to undergo a frequency shift as a function of the analytes concentration binding at the metasurface, defining the sensitivity of the sensing platform. 
To this end we use the experimental data of Koohyar et al.\cite{koohyar2012study} who tabulated the refractive index (RI) variations related to different molar fractions of  aqueous solutions of methanol, ethanol, ethylene glycol, 1-propanol and 1, 2, 3-propantriol. In particular, we consider the data corresponding to the  aqueous solution of 1, 2, 3-propantriol characterized by an ultra low molecular weight ($C_3H_8O_3 \approx$ 60 Da).

\begin{figure}
	\includegraphics[width=0.9\columnwidth]{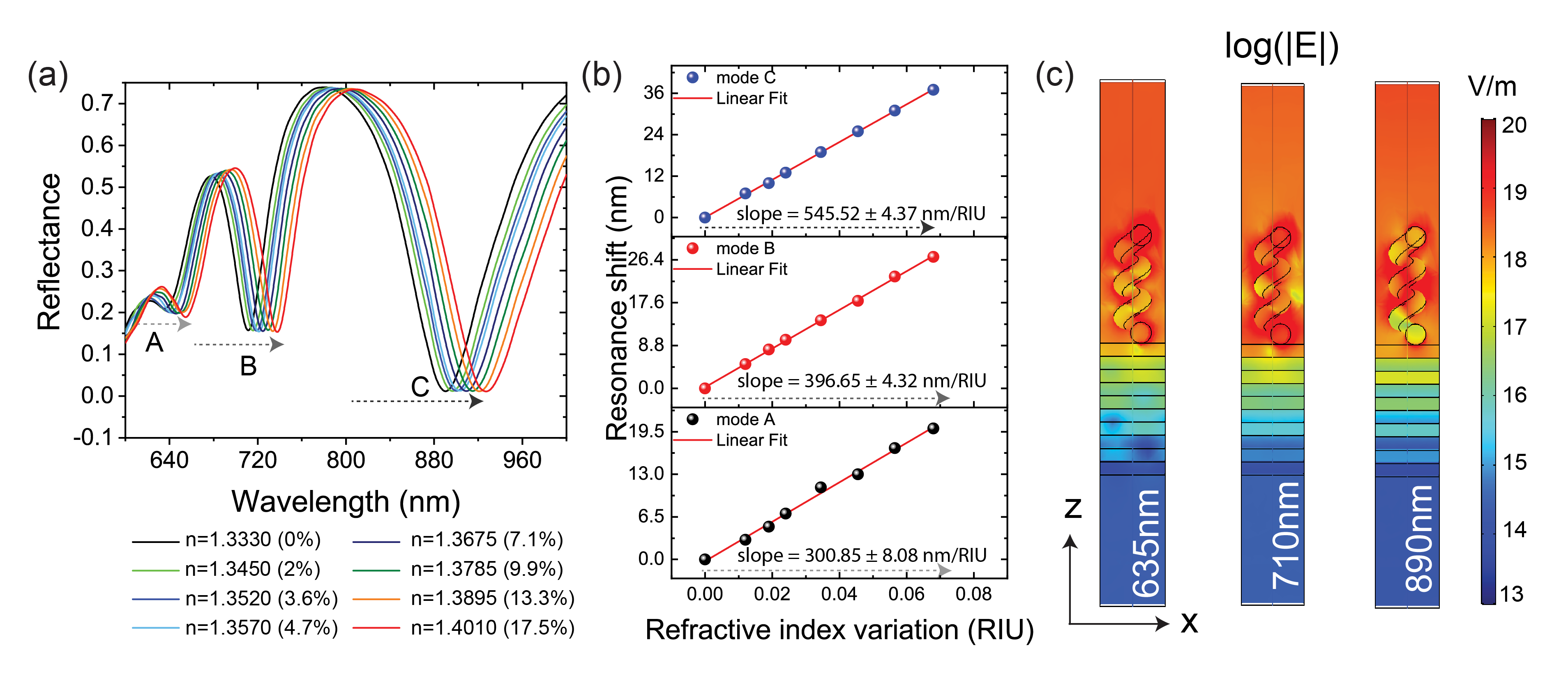}
	\caption{(a) Performance evaluation of the CMH-HMM sensor: calculated reflectance spectra  with different mole fractions of 1, 2, 3-propantriol  in distilled water. (b) Resonance wavelength shift for the mode A,  mode B and the third mode C as a function of the refractive index variation and corresponding linear fitting. (c)X-Z maps of the electric field intensity through the CMH-HMM structure in logarithmic scale for the three modes.}
	\label{fig:4}
\end{figure}

Figure \ref{fig:4}a shows the calculated reflection spectra of the CMH-HMM sensor in measuring 1, 2, 3-propantriol solutions with different concentrations.
The shift of the resonances as a function of the different percentage of propantiol in the aqueous solution allows us to determine the sensitivity of the sensor.
The reflectance minima (mode dips) shift towards longer wavelengths with the increase of RI, in the range $0 <\Delta_{RI} <0.0680$, this last case corresponding  to a molar fraction of $C_3H_8O_3$ of about 17\% in the aqueous solution; the resonance wavelength shift ($\Delta\lambda$) shows a linear behaviour with the RI change ($\Delta_{RI}$), as reported in Figure \ref{fig:4}b. 

The spectral sensitivity of the sensor is defined as $S= \Delta n/\Delta \lambda$, where $\Delta n$ represents the change of the refractive index solution flowing on top of the sensor and $\Delta\lambda$ the shift of the sensor resonance in $nm$.

In particular, $\Delta \lambda_A$=20 nm and $S_A$=3.3 10$^{-3}$ RIU/nm for the first mode A (635 nm), $\Delta \lambda_B$=26 nm and $S_B$=2.5 10$^{-3}$ RIU/nm for the second mode B (710 nm) and $\Delta \lambda_C$=37 nm and $S_C$=1.8 10$^{-3}$ RIU/nm for the third mode C (890 nm).  The corresponding limit of detection (LOD), corresponding to the minimum RI variation that can be distinguished (shift of 1 nm) results to be equal to 0.0015. 
The difference in sensitivity for the three modes is related to their different $\Delta \lambda$. In particular, mode C is the one that shows the best result in terms of sensitivity respect to the other two modes, this is due to the highly confined field distributions on the superstrate at hyperbolic dispersion (Figure \ref{fig:4}c) and to the fact that the transverse decay of the field in the HMM strongly varies from one mode to another.

Another important aspect related to the 3D hypergrating is represented by the significantly increase of the out-of-plane sensing surface, where the specific binding of the analytes occurs via functional groups immobilized in this specific case on a helical surface.  At the same time, a chiral structure can modify the fluid dynamics around it, inducing an increase of  the probability of a specific binding. 
 
 \begin{figure}
  \includegraphics[width=0.7\columnwidth]{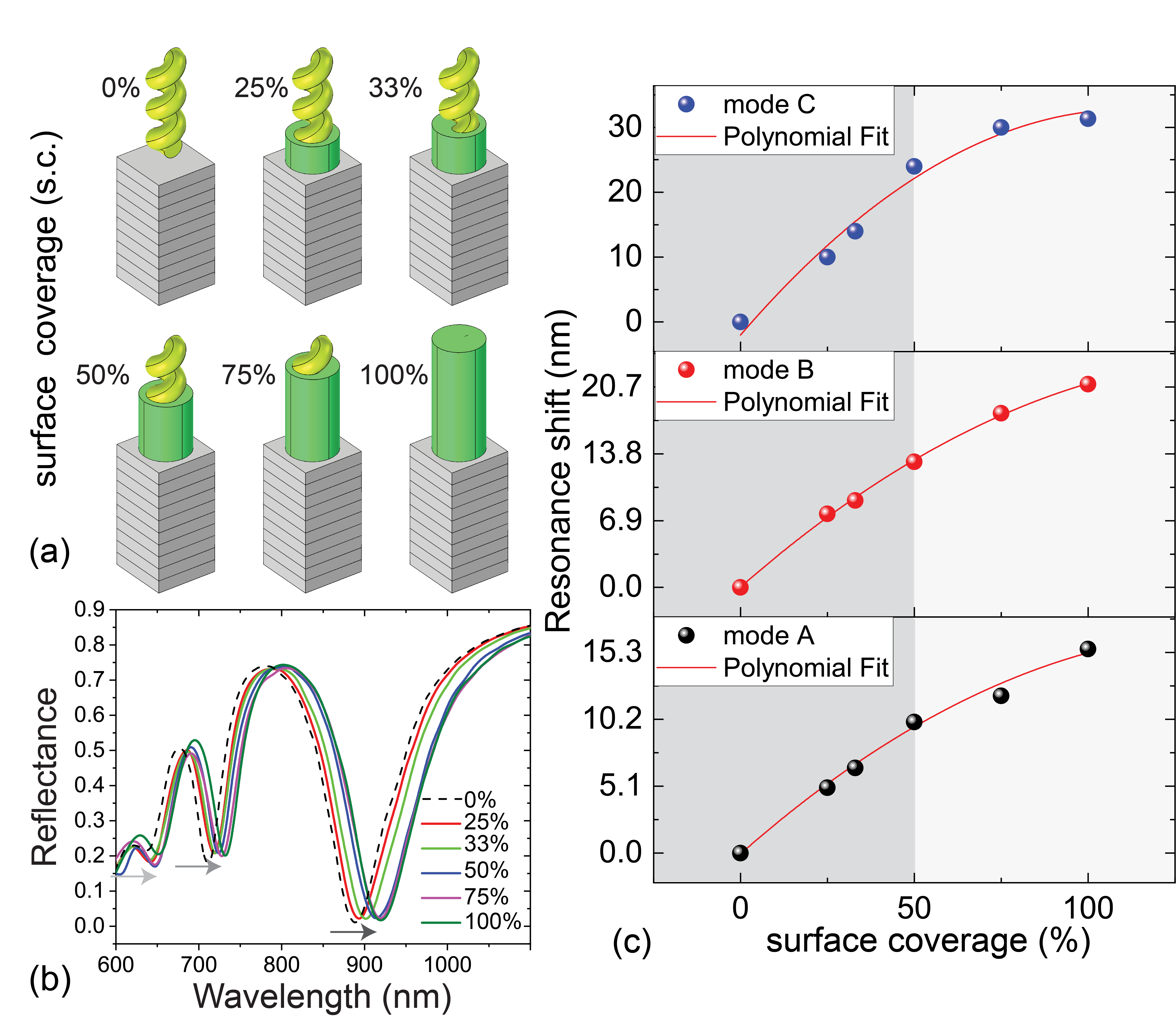}
  \caption{(a) Sketch of the simulated geometry with different percentage of the surface coverage and (b) corresponding reflectance spectra. (c) Resonance wavelength shift for the mode A, B and C as a function of the variation of the surface coverage. }
  \label{fig:5}
\end{figure}  

Clearly, the wavelength shift of BPP modes is strongly related to the quantity of molecules that bind selectively on the surface of the helices. 
To demonstrate this behavior  numerically, we have considered different percentages of surface coverage of the helicoidal structures, starting from the HMM surface (Figure \ref{fig:5}a), where the maximum variation of the considered refractive index ($\Delta_{RI}$=0.0680) affects only particular regions around the helix.
To this end, we have considered the two limiting cases: the 0\% of surface coverage (s.c.) and the 100\%, in which the whole surface of the helix is totally covered by molecules; in this last case we obtain the maximum  shift of 31 nm for mode C, 21 nm for mode B and 15 nm for mode A.
Then we considered four intermediate cases of partial s.c.: 25\%, 33\%, 50\%, and 75\% (Figure \ref{fig:5}b). 
As we can see in Figure \ref{fig:5}c the resonance wavelength shift of the helix/HMM modes shows a linear behaviour until 50\% of surface coverage, once this value is exceeded the slope of the curve decreases, by showing that the sensitivity is decreasing too.

At the same time, it is possible to find the minimum percentage value, that in our case results to be equal to 16\%,  corresponding to the minimum surface coverage, necessary to have an appreciable shift of the modes.
This leads to calculate the mass sensitivity, a key parameter used to describe the performance of a biosensor, that is strongly related to the surface coverage;
 for the considered system the mass sensitivity results equal to 0.04 $pg/mm^2$ (see Supplementary materials).

\begin{figure}[!h]
  \includegraphics[width=0.6\columnwidth]{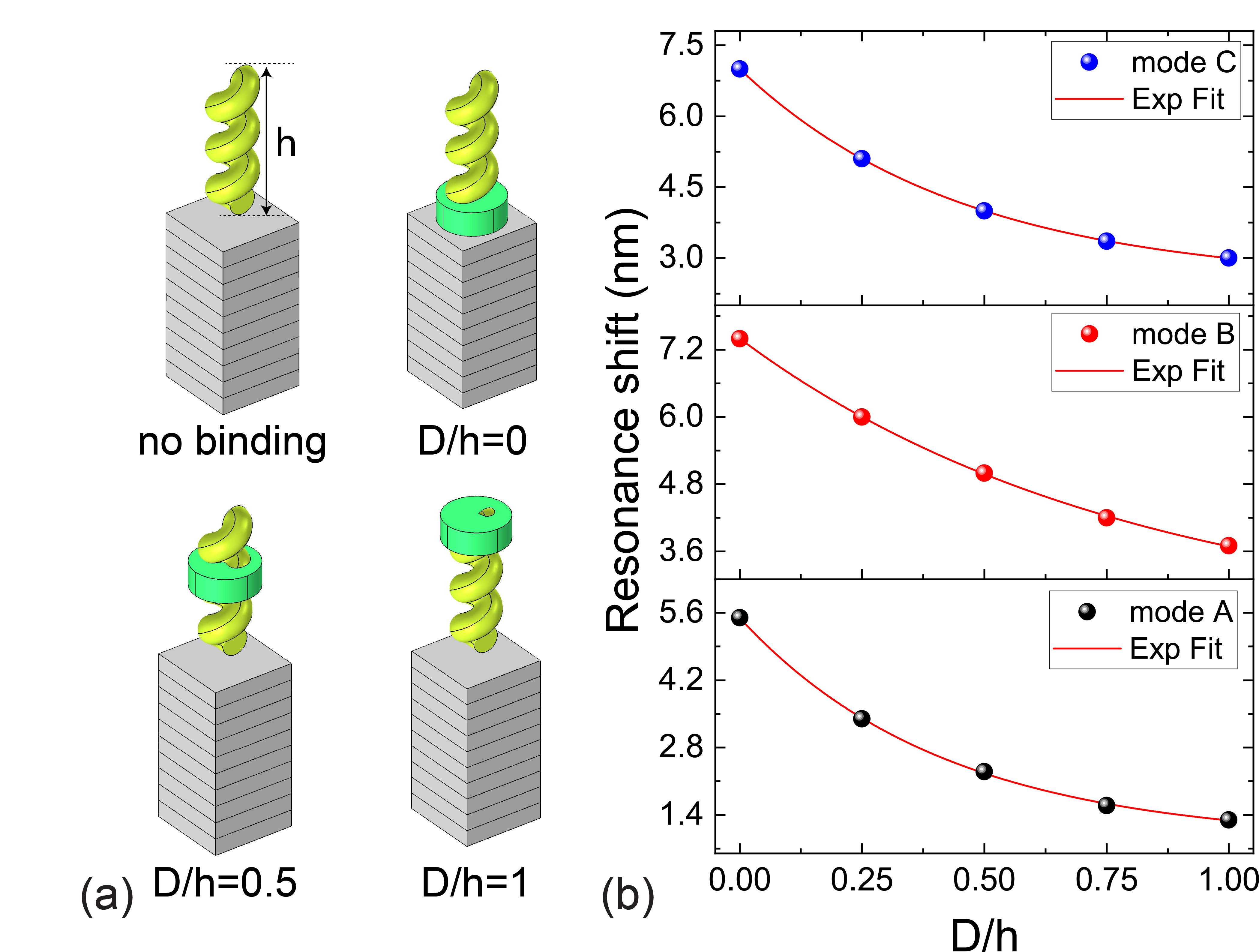}
  \caption{(a)Sketch of the  geometry with the surface coverage (s.c.) of 20\%  at two different heights. (b) Corresponding simulated reflectance spectra and (c) a magnification of them in proximity of the mode  A, B and C. }
  \label{fig:6}
\end{figure}

The last analysis is dedicated to investigate how the sensitivity changes as function of the relative surface coverage at different heights away from the HMM surface.  
In particular, it is possible to demonstrate that the local change of refractive index in a small disk surrounding the helix produces appreciable shifts even when the binding is confined exclusively in the upper region of the helix, which represents the maximum distance from the HMM.
For this reason a small disk, with $n$=1.401 and corresponding to an adsorbed surface of 20\%, is positioned at different distance ($D$) respect to the surface of the HMM (Figure \ref{fig:6}a).  By plotting the resonance shift of the three modes A, B and C as a function of the distance $D$ normalized to the helix height ($h$) it is possible to see that the proposed biosensing platform is able to appreciate a shift of the considered modes even in the worst case ($D/h$=1) of about 1.4 nm for the mode A, 3.6 nm for the mode B and 3.0 nm for the mode C, as reported in Figure 6b. 
It is interesting to note that the resonance shift decreases exponentially by moving away from the surface of the HMM.
The corresponding sensitivity results to be 48.6 10$^{-3}$RIU/nm for the mode A, 18.9 10$^{-3}$RIU/nm for the mode B and 22.6 10$^{-3}$RIU/nm for the mode C. 
These results demonstrate the strong sensing power of our innovative system to detect target analytes also away from the surface of the HMM exploiting the increased surface/volume ratio of the 3D CMH exposed to the analyte.

\begin{figure}
	\includegraphics[width=0.8\columnwidth]{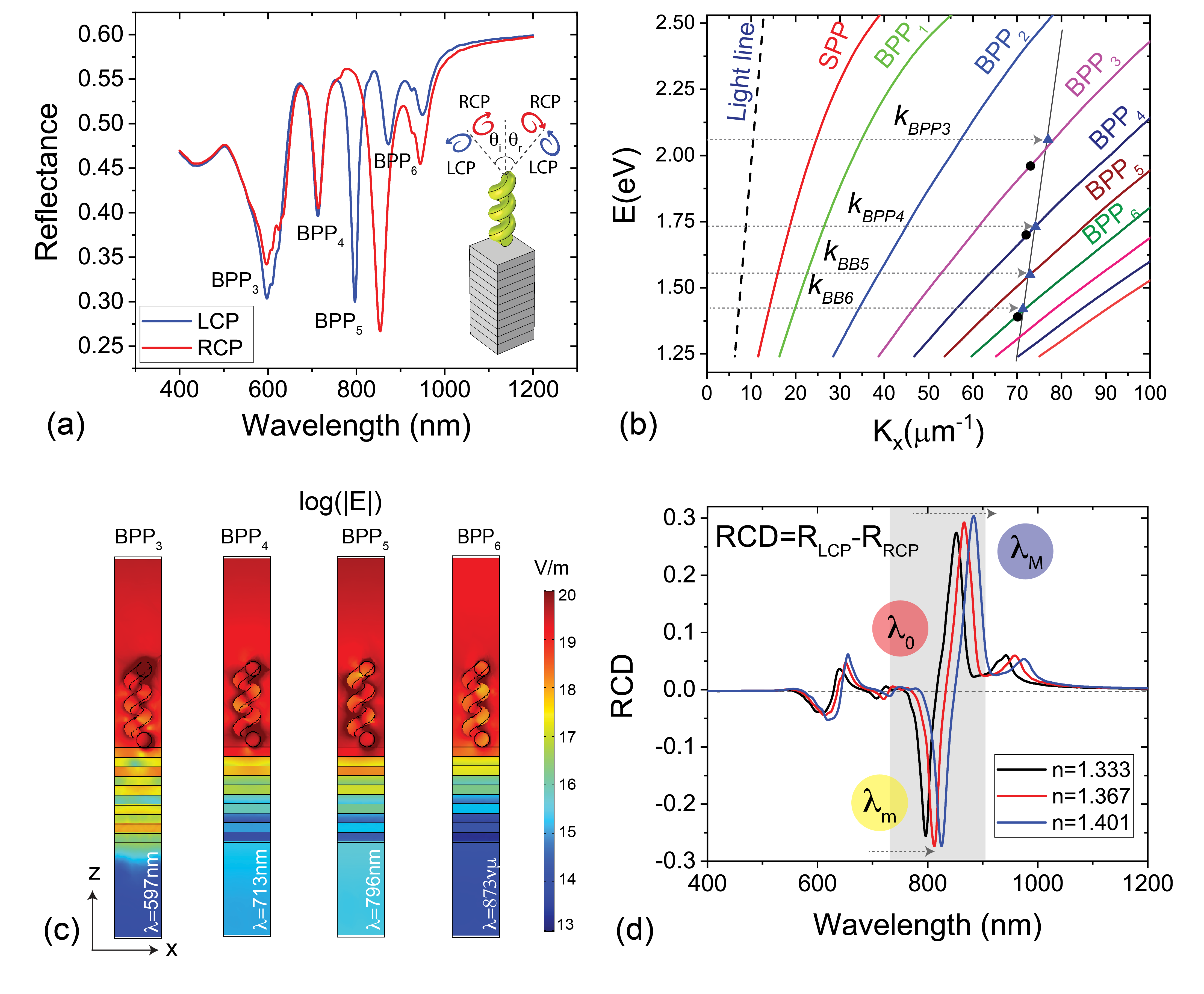}
	\caption{(a)Reflectance curves of the CMH-HMM for LCP and RCP light, angle of incidence $\theta_i$=75$^\circ$; in the inset a sketch of the simulated unit cell probed with circular polarized light. (b) Intersection  of the dispersion curves of the HMM with BPP modes excited by circular polarized light. (c) X-Z maps of the electric field intensity through the CMH-HMM structure in logarithmic scale for LCP light and $\theta_i$=75$^\circ$. (d) Reflectance circular dichroism (RCD) as a function of refractive index variations.}
	\label{fig:7}
\end{figure}

On the other hand, considering the intrinsic chirality of the nanohelices, CMH could excite new BPP modes of the HMM by coupling with their circular polarization-dependent plasmon modes. Indeed, as we can see in Figure \ref{fig:7}a, for an angle of incidence of $\theta_i$=75$^\circ$, different reflectance dips are obtained for left-handed circular polarized (LCP) and right-handed circular polarized (RCP) light. In particular, the reflectance dips obtained for LCP light, indicated in the figure as BPP$_3$ to BPP$_6$, are strongly related to a coupling between the plasmonic modes of the gold helix array and the HMM underneath. This can be verified through the intersection of the dispersion curves of the HMM with the wavelength corresponding to the reflectance dips that are expressed in terms of energy versus momentum and are indicated in the Figure \ref{fig:7}b with the blue triangles. For LCP light, at these wavelengths, respectively: 597 nm, 713nm, 796 nm and 873 nm the electric field intensity |E| results to be confined between the helix antenna surface and the HMM layers - Figure \ref{fig:7}c. A similar analysis could be carried out for RCP reflectance dips.

From the analysis of the chiroptical response of the CMH we can calculate the CD signal, typically riches of spectral signatures, such as bipolar peaks and crossing points, increasing the sensitivity and accuracy of the  monitoring refractive index changes due to the analyte absorption. This sensing modality offers strong optical contrast even in the presence of highly achiral absorbing media, increasing the signal-to-noise CD measurements of a chiral analyte, an important consideration for use in complex biological media with limited transmission.\cite{jeong2016dispersion}
For this purpose, the resulting reflectance curves obtained for LCP and RCP light, at $\theta_i$=75$^\circ$ are used to calculate the reflectance circular dichroism (RCD) spectrum, which characterizes the reflectance difference between LCP and RCP light, leading to a RCD amplitude (RCD=R$_{LCP}$-R$_{RCP}$). 
As expected, the RCD spectra for the CMH-HMM exhibit multiple features: different maxima, minima and crossing points (Figure \ref{fig:7}d).
We have studied the RCD, as a function of the refractive indices variation between 1.333 and 1.401 (glycerol - water mixtures varying from 0 to 17.5\% concentration - Figure \ref{fig:7}d) to quantify the performance in terms of sensitivity and FOM of our sensor. 

We can distinguish in the range 700-900 nm a minimum ($\lambda_m$), a crossing point ($\lambda_0$) and a maximum ($\lambda_M$) respectively at 797 nm, 816 nm and 852 nm which significantly shift and modify their intensity as the refractive index changes. 
These signals show a chiral plasmon sensitivity S$_{CP}$=$\Delta n/\Delta\lambda$ of 2.42 10$^{-3}$RIU/nm for $\lambda_m$, 2.12 10$^{-3}$RIU/nm for $\lambda_0$ and 2.19 10$^{-3}$RIU/nm for $\lambda_M$, respectively. After extracting the classical FWHM for the $\lambda_m$ and $\lambda_M$ modes and the FWHM in the |RCD| spectrum (see Supplementary materials) for $\lambda_0$, we have calculated the FOM,\cite{mayer2011localized} defined in our case as $FOM$=$(S \cdot FWHM)^{-1}$, that resulted to be, respectively, equal to 18.72, 15.18 and 15.19, very high values if compared to previously reported results.


\section{Conclusions}

In this paper, we have numerically demonstrated how an out-of-plane  chiral metasurface  hypergrating can be used to excite the plasmonic modes of an underlying hyperbolic metamaterial with linearly and circularly polarized light. The electromagnetic modes resulting from the coupling between the CMH-HMM depend strongly on the material and geometrical parameters of the CMH-HMM. In our numerical study, we have demonstrated how the HMM BPPs modes can be used for the detection of ultra low molecular weight molecules ($\approx$ 60 Da) with extremely low concentrations (mass sensitivity of 0.04 $ pg / mm ^ 2 $) on the whole surface of the helices array. 
The CMH represents a new versatile way to couple circularly polarized light with the HMM, by exploiting the features of the reflectance circular dichroism (RCD) spectra, rich of signals that can be used to increase the processed data for the detection of analytes by the proposed biosensing platform.

\begin{acknowledgement}
This research has been supported by the \textit{"AIM: Attraction and International Mobility"} - PON R\&I 2014-2020 Calabria and  "Progetto Tecnopolo per la Medicina di precisione, Deliberazione della Giunta Regionale n. 2117" Puglia.  The authors thank the Area della Ricerca di Roma 2, Tor Vergata, for the access to the ICT Services (ARToV-CNR) for the use of the COMSOL Multiphysics Platform, Origin Lab and
Matlab, and the Infrastructure BeyondNano (PONa3-00362) of CNR-Nanotec for the access to research instruments.
\end{acknowledgement}

\bibliography{Helix_sensing.bib}
\bibliography{Helix_sensing.bbl}

\begin{suppinfo}

Contents:

\begin{itemize}
  \item Electric permittivity of the HMM calculated by means of the effective medium theory 
  \item Equations for Scattering parameters $S_{11}$ and $S_{21}$ 
  \item Meshed geometry
  \item Unchiral hypergratings
  \item Calculation of the mass sensitivity
  \item  Detailed plots for the FOM calculation
\end{itemize}

\end{suppinfo}


\end{document}


Contents:

\begin{itemize}
 \item Electric permittivity of the HMM calculated by means of the effective medium theory 
 \item Meshed geometry
 \item Equations for scattering parameters $S_{11}$ and $S_{21}$ 
  \item Unchiral hypergratings
 \item Calculation of the mass sensitivity.
 \item  Detailed plots for the FOM calculation.
\end{itemize}

\begin{itemize}

\item Electric permittivity of the HMM calculated by means of the effective medium theory  based on Maxwell-Garnett theory:
 
 Effective Medium Theory (EMT) of the ITO/Ag multilayer. 
 The optical constants, $n$ and $k$, used in the numerical simulation for Silver were obtained from the Johnson and Christy (1972) 0.1888 - 1.973 $\mu$m contained within the Material Library in COMSOL Multiphysics.
 For the In$_2$O$_3$-SnO$_2$ (ITO, Indium tin oxide) were considered the numerical data from K\"onig et al. (2014)  from the Material Library in COMSOL Multiphysics too.

\begin{figure}
  \includegraphics[width=0.5\columnwidth]{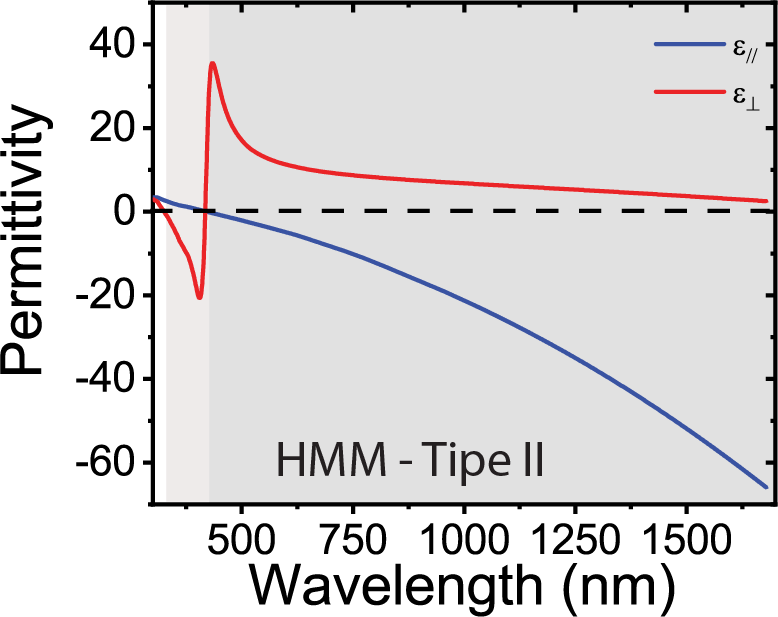}
  \caption{Real part of $\epsilon_{//}$ (blue curve) and $\epsilon_{\perp}$  (red) of the HMM structure.}
  \label{fig:S1}
\end{figure}

\setlength{\parskip}{1em}


\item Meshed geometry:

An important role in the numerical simulation is carried out by the mesh that discretizes the problem. COMSOL Multiphysics gives the opportunity to set it as controlled by the physics, by selecting the option "physics-controlled the mesh" present in "physics study". A "normal mesh" is automatically generated by the software, depending on the minimal size present in the system, compared with the incident wavelength.
A sketch of the mesh directly controlled from the physics is reported in Figure \ref{fig:S2}, the larger part are meshed with a normal weave (superstrate, glass) while for the small parts (helix, HMM)  a fine mesh weave has been used.

\begin{figure}[!ht]
  \includegraphics[width=0.6\columnwidth]{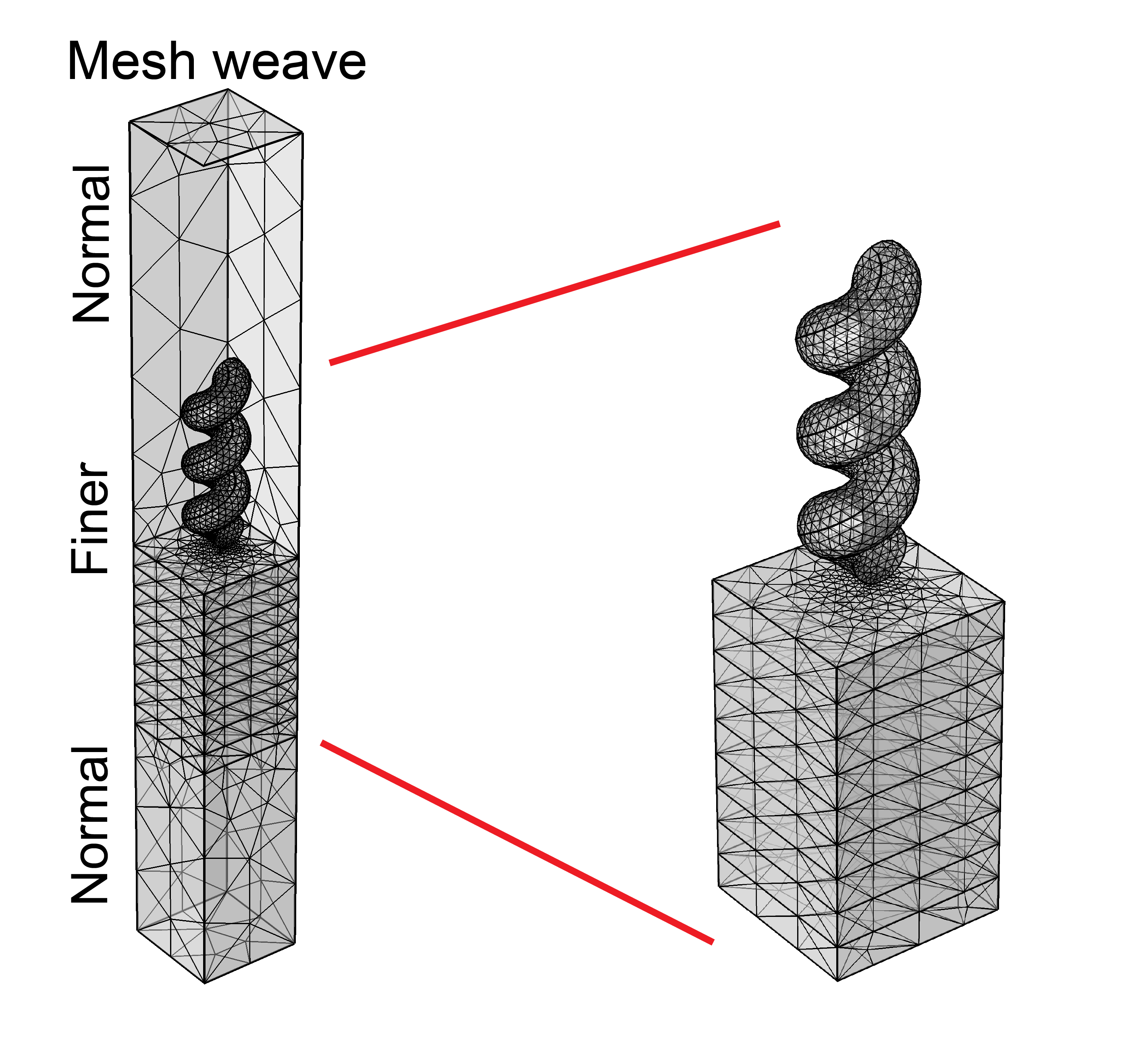}
  \caption{A sketch of the mesh directly controlled from the "physics-controlled mesh" tool of COMSOL.}
  \label{fig:S2}
\end{figure}

\setlength{\parskip}{1em}


\item Equations for scattering parameters $S_{11}$ and $S_{21}$:

The S-parameter for the mode with index $k$ is given by multiplying with the conjugate of the mode field for mode $k$ and integrating over the port boundary. Since the mode fields for the different modes are orthogonal, the following relations are obtained for the S-parameters:

\begin{equation}
S_{11}=\frac{\int\limits_{port1}((\mathbf{E_c}-\mathbf{E_1}) \cdot \mathbf{E_1}^* )dA_1}{\int\limits_{port1}(\mathbf{E_1} \cdot \mathbf{E_1}^*)dA_1}
\label{eq1}
\end{equation}

\begin{equation}
S_{21}=\frac{\int\limits_{port2}((\mathbf{E_c}-\mathbf{E_2}) \cdot \mathbf{E_2}^* )dA_2}{\int\limits_{port2}(\mathbf{E_2} \cdot \mathbf{E_2}^*)dA_2}
\label{eq1}
\end{equation}

where $E_c$ is the computed electric field equal to $ \mathbf{E_c}=\sum\limits_{i=1}  S_{i1}\mathbf{E_i}$, and by assuming that the fields are normalized with respect to the integral of the power flow across each port cross section $A_1$, respectively.

\setlength{\parskip}{1em}

\item Unchiral hypergratings:
 
 \begin{figure}
 	\includegraphics[width=0.6\columnwidth]{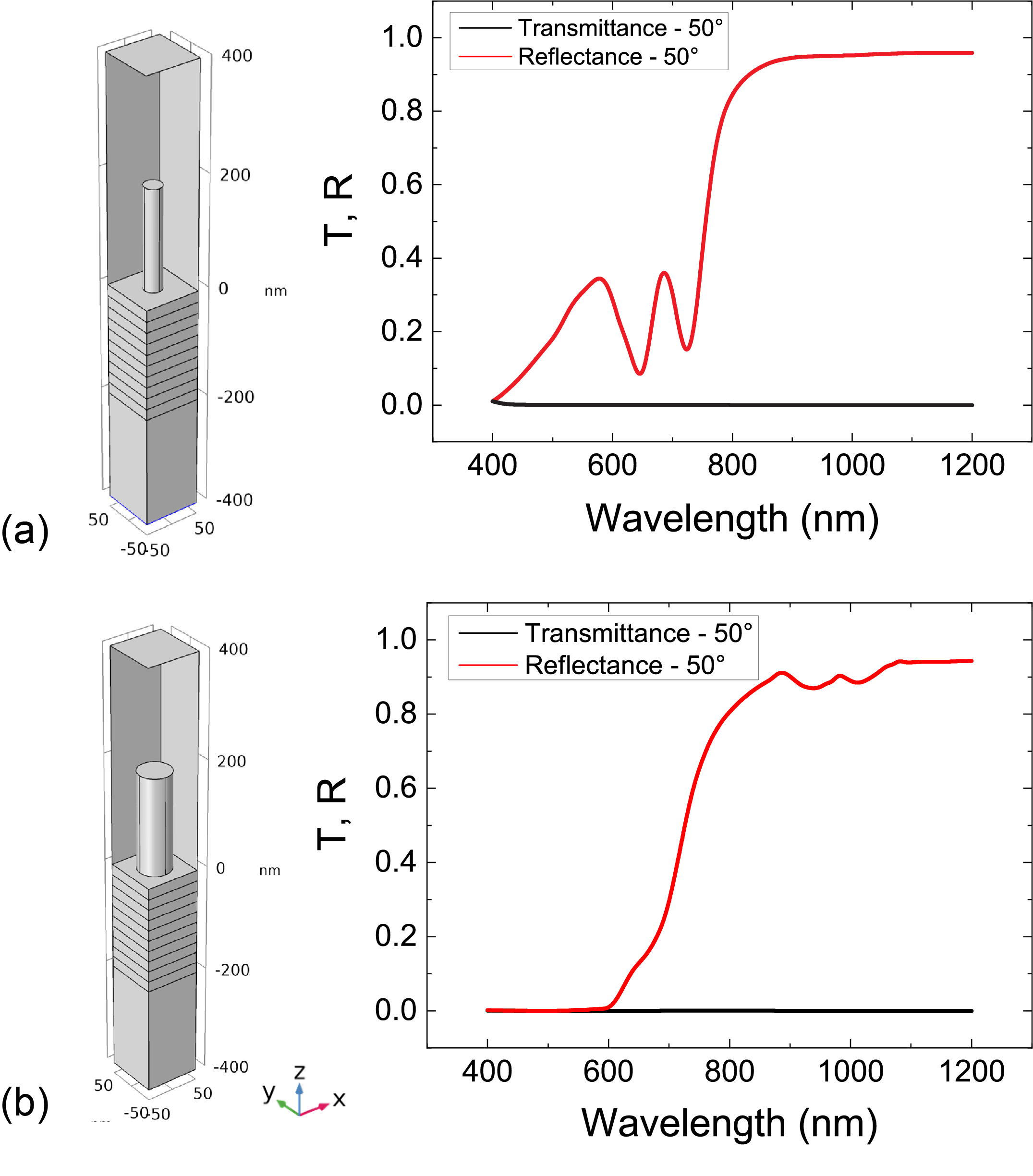}
 	\caption{Sketch of the unit cell of pillar hypergrating characterized by the same radius, $R_h$, (a-left) and  wire radius $r$ (b-left) of the helix with corresponding transmittance and reflectance curves (a-b right) for the two considered geometry.}
 	\label{fig:S3}
 \end{figure}
 
 The reflectance and transmittance curves for an out-of-plane pillar hypergrating  characterized by the same sizes of the wire radius ($r$) and of the helix radius ($R_h$)  - Figure \ref{fig:S3}  - are reported. As we can see the unchiral hypergratings, with the same material parameters considered for the helix hypergrating do not shown any interesting signals in reflectance and transmittance. The two curves are calculated by considering a TM light, with an angle of incidence of 50$^\circ$. 
 

\item Calculation of the mass sensitivity:

The minimum surface coverage of the helix, that leads to an appreciable shift  of the plasmonic modes, correspond to 16\% ($S'$)  of its total surface. This value is  approximately equal to 5144 nm$^2$. The hydrodynamics radius of the C$_3$H$_8$O$_3$ molecules is approximately $r_H \approx 0.3 nm$, with a molecular weight of about 60 Da. In $S'$ we can find approximately ($S'/(\pi r^2)$) numbers of  C$_3$H$_8$O$_3$  molecules that corresponds to 18195 molecules. 
The weight in $pg$ of a single molecule of C$_3$H$_8$O$_3$ corresponds to: 60 x 1.66 x 10$^{-18} \mu$g $\approx 10^{-10}$ pg that results to be 18.2 x 10$^{-7}$ pg for the total number of considered molecules.
By dividing this number for the S' we can obtain the mass sensitivity  of our system ($\approx 0.04 pg/mm^2$) that represents the minimum quantity of absorption molecules that can be appreciate in terms of plasmonic modes shift. 


\item Detailed plots for the FOM calculation:

Detailed plots of the resonance shifts at  $\lambda_m$ and $\lambda_M$, respectively. Calculation of |RCD| for the FWHM related to $\lambda_0$. 

 \begin{figure}
 	\includegraphics[width=0.9\columnwidth]{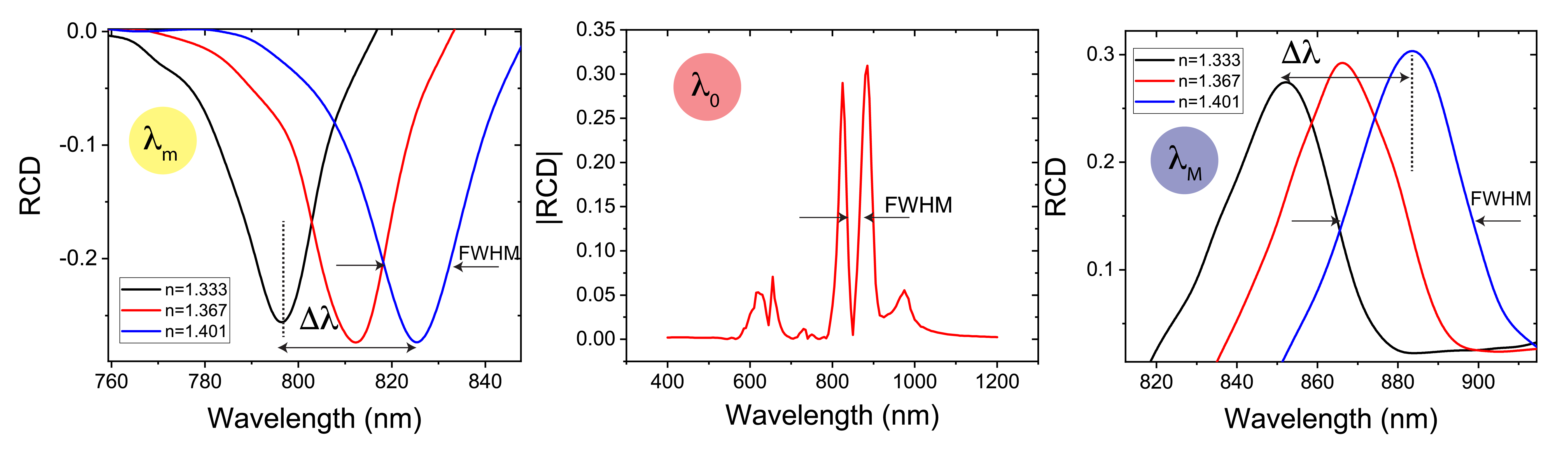}
 	\caption{Resonance shift for$\lambda_m$ and $\lambda_M$ and |RCD| calculation for the FWHM of $\lambda_0$.}
 \end{figure}

\end{itemize}
 